\documentclass[english]{aa}
\usepackage{graphicx}
\usepackage{natbib}
\bibpunct{(}{)}{;}{a}{}{,} 
\usepackage{txfonts}
\usepackage[latin1]{inputenc}
\titlerunning{Driving Mechanism in Abell 43}

\begin{document}
   \title{The Nature of the Driving Mechanism in the Pulsating Hybrid PG
     1159 Star Abell~43}

   \author{P.-O. Quirion,
            G. Fontaine, and P. Brassard }

   \offprints{P.-O. Quirion}

   \institute{Département de Physique, Université de Montréal,
               C.P. 6128, Succ. Centre-Ville, 
               Montréal, Québec H3C 3J7, Canada\\
              \email{quirion@astro.umontreal.ca,
               fontaine@astro.umontreal.ca, brassard@astro.umontreal.ca}
             }

   \date{Received; accepted}

   \abstract{
   We extend our previous pulsational stability analyses of PG~1159
   stars by modeling the hybrid~PG~1159 type star Abell~43. We show that the
   standard $\kappa$-mechanism due to the ionization of C and O in the
   envelope of this H-rich PG~1159 star is perfectly able to drive
   g-mode pulsations. Thus, contrary to a recent suggestion, there is no
   need to invoke any new or exotic mechanism to explain the pulsational
   instabilities observed in this particular star. Our expected
   instability band for $l=1$ modes extends in period from $\sim 2604$ s
   to $\sim 5529$ s, which is consistent with the available photometric
   observations of Abell~43. We also suggest that efforts to detect
   luminosity variations in its sibling NGC 7094 be pursued.  

   }

   \maketitle
\section{Astrophysical Context}

	Abell~43 is, along with NGC~7094, HS~2324+3944 and Sh~2-68, one of
        the four known hybrid PG~1159 stars. PG~1159 stars are usually
        observed as hydrogen deficient stars, but the so-called hybrid
        PG~1159 objects reveal strong H Balmer lines indicative of a
        non-negligible hydrogen abundance in their atmospheres 
        \citep{NapScho91}. The latest published atmospheric parameters
        for Abell~43 are those of \citet{Mik02} where we find that
        $T_{\rm eff} = 110,000$ K, $\log g = 5.7$, $X$(H)=0.35,
        $X$(He)=0.42, and $X$(C)=0.23, the latter in units of mass
        fraction. Abell 43 is surrounded by a planetary nebula and is
        classified as a lgEH object in the \citet{Werner92} notation. In
        this paper, the term PG~1159 will include hybrid PG~1159 stars.

	\citet{Herwig01} first showed that both partial and total depletion
        of hydrogen in PG~1159 type stars can be explained if the last
        thermal pulse of the post-AGB evolutionary phase appears either 
        early (the so-called AGB Final Thermal Pulse or AFTP) or late 
        (the Late [LTP] or Very Late Thermal Pulse [VLTP]). An
        interesting characteristic of stars that undergo AFTP evolution 
        is that they do not pass through any Born Again phase. The last
        He~flash occurs just as they start to heat up along the post-AGB
        track, long before the white dwarf cooling phase. In these
        circumstances, the convection zone provoked by the He~flash does
        not reach the surface, so the hydrogen present in the star is not 
        completely burned away. It is the third dredge-up, happening
        later, that will mix the surface and leave it with some finite 
        amount of H. The \citet{Herwig01} paper presents two AFTP
        models ending up as PG~1159 stars with surface mass fractions of
        $X$(H)=0.17, $X$(He)=0.33, $X$(C)=0.32, $X$(O)=0.15 and
        $X$(H)=0.55, $X$(He)=0.31, $X$(C)=0.07, $X$(O)=0.04, thus giving 
        great credibility to this formation scenario for hybrid PG 1159
        stars.

	Interest in Abell 43 -- a peculiar object of a peculiar class -- is
	  compounded by the fact that it shows multiperiodic luminosity
	  variations. The possible variability of Abell~43, with a main period of
	  about 2473 s, was first reported by \citet{CiaBon96}.  These authors,
	  however, used very conservative stability criteria and hesitated to
	  conclude that Abell~43 is most likely a variable star. For their part,
	  \citet{Schu00} reported that the  5500 s period they observed was a
	  real detection at the 99\% confidence level, and considered that this
	  variation, the longest ever observed in a PG 1159 star, had a chance of
	  being the orbital period of a binary system. More recently,
	  \citet{Vau05} positively observed 2600 s and 3035 s luminosity
	  variations in Abell~43 which they assigned to $g$-mode pulsations.
	  However, they doubt the fact that those modes could be driven via the
	  classic $\kappa$-mechanism due to carbon K-shell electrons ionization
	  as in all other known pulsating PG 1159 stars (see, e.g., Quirion et
	  al. 2004a). Their doubts were brought about by the use of the
	  \citet{Dre95} table, where preliminary atmosphere modeling of Abell~43
	  gave mass fraction ratios of  $X$(H)=0.42, $X$(He)=0.51, $X$(C)=0.05,
	  with $\log g = 5.7$ and $T_{\rm eff} = 110,000$ K. This is a very low
	  carbon abundance and it, indeed, induces no driving when modeled. We
	  show in this paper, by using the  revised table published by
	  \citet{Mik02} that doubts about the nature of the driving mechanism in
	  Abell~43 are unfounded.

	Early stability calculations \citep{Star83,Star84,Star85,Stan91}
suggested that the C/O  $\kappa$-mechanism in PG~1159 stars was
suffering from the so-called helium and hydrogen poisoning phenomenon. 
Indeed, it was found in those computations that the composition in the
driving region had to be highly helium- and hydrogen-poor in order to be
able to drive pulsation modes. The hydrogen poisoning, contrary to
helium poisoning, was not considered as a real problem then since 
pulsating hybrid PG~1159 stars were not known at the time. However, the
helium poisoning problem posed a real challenge to theory because it
would require the existence of very important composition gradients
between the surface (typically rich in helium in PG 1159 stars) and the
driving region located deeper in the envelope. No credible explanation
was ever put forward to account for such large composition gradients
and, furthermore, the whole concept was in direct conflict with the idea
(and subsequent observational evidence) that residual stellar winds
pervade the outer layers of PG 1159 stars and tend to produce homogeneous
envelopes.

        The puzzle of the helium poisoning problem was solved by 
\citet{Saio96} who convincingly showed that the problem was related to 
inadequacies in the older Los Alamos opacity data used in the previous 
stability calculations. \citet{Saio96} demonstrated that the problem
simply went away with the use of the then newly available OPAL tables. 
In the meantime, a first hybrid PG~1159 pulsator, HS~2324+3944, showing
a substantial atmospheric abundance of hydrogen ($X$(H) $\sim$ 0.2) was
discovered by \citet{Silvotti96}. \citet{Gautschy97} was then able to 
produce a satisfactory instability strip for models resembling 
HS~2324+3944 including up to $20\%$ (mass fraction) of hydrogen in the
driving region. His calculations, again based on the OPAL opacity data,
revoked once and for all the former allegations of hydrogen poisoning.
Finally, \citet{QFB04} underwent a larger model survey of PG~1159 stars,
first reproducing Saio's and Gautschy's results, and then explaining the
coexistence of variable and nonvariable PG~1159 stars in the $\log
g$-$T_{\rm eff}$ diagram in terms of a dispersion in atmospheric
parameters and in terms of a variation in surface composition from star to
star. This last study (also based on the OPAL dataset) confirms that the
$\kappa$-mechanism due to C/O ionization (and working in an envelope with
a uniform chemical composition) is the only mechanism needed to explain the
instability properties of PG~1159 stars. We complete here the survey by
modeling in the same terms as before the hybrid PG~1159 star Abell 43,
the one with the highest known atmospheric hydrogen abundance. Contrary
to the suggestion of \citet{Vau05}, we find that there is no need to
invoke a new or different excitation mechanism in this particular object.

\section{Models}

To model Abell~43, we used the same approach as that presented in
\citet{QFB04}. We assumed a homogenous composition consistent with the
atmospheric composition determined by \citet{Mik02} down to a fractional
mass depth of $\log q = \log (1-M(r)/M_{*}) = -2.9$. For that part of
the equilibrium model (the envelope), we thus considered a mixture made
of (in mass fractions) $X$(H)=0.35, $X$(He)=0.42, $X$(C)=0.22, and $X$(Z)=0.01.
It should be mentioned that the assumed metallicity is in solar proportions,
except for C and O which are completely absent from the metal mixture
[$X$(Z)]. Hence, our envelope is completely free of oxygen and has an
exact carbon mass fraction of 0.22. These details have some importance
because small variations in the abundance of one of these elements can
change the efficiency of the $\kappa$-mechanism, especially when a star
lies near the edge of the instability strip. Note that the corresponding
opacity table has been ordered on the OPAL website, which is a most
useful facility. As indicated above, the effective temperature and the
gravity that we used for Abell 43 are $T_{\rm eff}$= 110,000 K and $\log g=
5.7$. The mass of the object, $M_{\star}=0.6~M_{\sun}$, has been chosen
to be representative of a typical PG 1159 white dwarf. It has already
been shown \citep{QFByale, QFB04} that the total mass has only a small
influence on the location of the instability range of PG~1159 models
with similar atmospheric parameters.

\begin{figure}
   \centering
   \includegraphics[angle= 90, width= 12 cm]{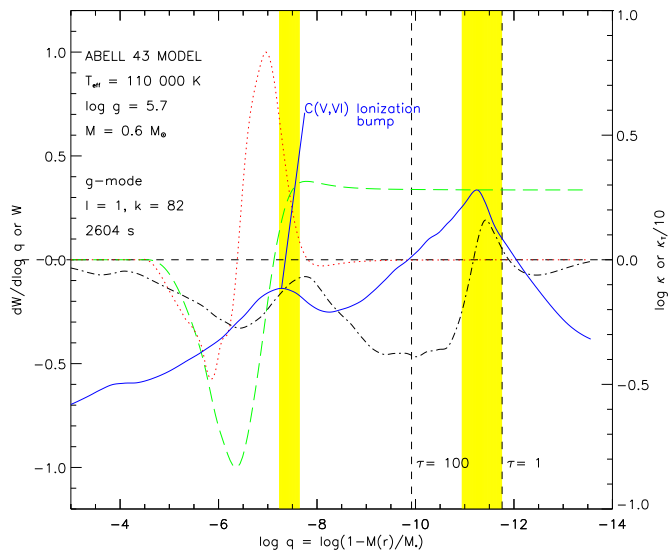}
      \caption{ Details of the driving/damping process in our model of
        Abell~43 for a typical excited $g$-mode. The figure shows the
        envelope, the part of the stellar model where the chemical
        composition is uniform. The location of the photosphere,
        $\tau_{Ross} \sim 1 $, is indicated by a vertical dashed
        line, while the position of the base of the atmosphere,
        $\tau_{Ross}$ = 100, is indicated by a similar line. The
        yellow stripes give the locations and extents of two 
        subphotospheric convection zones. The first one is just below
        $\tau_{Ross}=1$, and carries practically no flux, less than one
        tenth of a percent of the total flux. The second convection zone is
        linked to the C ionization bump and is also carrying little
        flux, less than $1.5\%$. The red dotted curve refers to the derivative
        of the work integral $dW/d$log $q$, and the green dashed curve to the
        running (from left to right) 
        work integral $W$. Both quantities are normalized to an
        extremum value of either +1 or $-$1. Also plotted is the run
        of the Rosseland opacity (blue solid curve) and the run of its
        logarithmic derivative $\kappa_T$ (black dot-dashed curve). The mode of
        interest has a period of 2604 s, an e-folding time of
        $\sim 1300$ yr, and a growth rate of $\sim 1 \times 10^{-8}$.
              }
         \label{fig:abell43}
   \end{figure}

The numerical tools for investigating the stability of our model were
the same as those used in \citet{QFB04}. We carried out our search for
unstable $l=1$ $g$-modes with periods from 1000 s to 6000 s and
found instability in the range from 2604 s to 5529 s. This covers
rather well the range of periods detected by various authors in Abell 43
as mentioned above. We show in Figure \ref{fig:abell43} some important 
characteristics of our Abell~43 model. The transition from the
homogenous envelope to the C/O core of the model has been put at 
$\log q = -2.9$, deep enough to have no significant influence on the
driving/damping region located well above $\log q \simeq -4$. The only
influence this deep transition zone could have on the stability analysis
is a small ``accordion effect'' on the boundaries of the instability
range as described in \citet{QFB04}. Essentially all the driving and
damping is confined to a region of the stellar envelope associated with
the bump in the opacity profile caused by CV and CVI ionization. This is
well illustrated by the red curve in Figure \ref{fig:abell43} which shows the
typical profile of the derivative of the work integral, $dW/d\log q$, for
a mode driven by the $\kappa$-mechanism. Our present results, along with
our previous ones, demonstrate beyond any doubt that the standard
$\kappa$-mechanism associated with C (and/or O) ionization in modern
models of PG 1159 stars is sufficient to explain the pulsational
instabilities seen in these stars, even those of the H-rich, hybrid
spectral type.

We extended, in a natural way, this study to NGC~7094, the last
hybrid PG~1159 star with known and reliable atmospheric parameters which
has not been modeled yet. According to Miksa et al. (2002), NGC~7094 has the
same parameters as Abell~43 except for a slight variation in atmospheric
chemical composition: $1\%$ less He by mass that is replaced by $1\%$
O. Knowing that the K-shell ionizations of C and O take place essentially
in the same region of our stellar model, it was reasonable to expect that
the driving mechanism would still be present with only very minor changes. 
And indeed, we found that, in our NGC~7094 model, pulsational
instabilities of $l=1$ $g$-modes are present in the period range
2550$-$5413 s. Thus, the effect of adding some trace of oxygen is to
shift the band of unstable periods to slightly lower values compared to
Abell 43, but the differences are quite small, as expected. In the light
of this result, we strongly suggest that the 1996 efforts, undergone by
\citet*{CiaBon96} to detect pulsations in NGC~7094, be continued.

\section{Conclusion}

Being peculiar is the norm for the PG~1159 stars. They differ in
atmospheric chemical composition, they cover a vast region in effective
temperature-surface gravity space, some have a planetary nebula, and some
show pulsational instabilities while others do not. Their location in
the H-R diagram is one of the few places where we expect similar stars
to have different looks. However, those differences in look should not
overshadow the fact that those stars are all subdued to the
same mechanisms. In the specific case of the pulsating PG~1159 stars,
there is an unifying theme, and that is the fact that all of them excite
pulsation modes through the same $\kappa$-mechanism associated with the
partial ionization of C and/or O in their envelope. There is no need to
invoke any new or exotic driving mechanim, even for those H-rich
pulsating PG~1159 stars of the hybrid type.

This study completes the work of \citet{QFB04} in that we have now
modeled all PG~1159 stars with available reliable values of their
atmospheric parameters and studied their pulsational stability. We have
offered a consistent account of these stars, including the nonpulsators
as well. We look forward to the results of future searches for
pulsations in the hybrid PG~1159 star NGC~7094.

\begin{acknowledgements}
This work was supported in part by the Natural Sciences and Engineering
Research Council of Canada and in part by the Fonds Qu\'ebecois de la
recherche sur la nature et les technologies (Qu\'ebec). G.F. also
acknowledges the contribution of the Canada Research Chair Program.
\end{acknowledgements}

\bibliographystyle{aa} 
\bibliography{bibliographie} 

\begin{thebibliography}{17}
\expandafter\ifx\csname natexlab\endcsname\relax\def\natexlab#1{#1}\fi

\bibitem[{{Ciardullo} \& {Bond}(1996)}]{CiaBon96}
{Ciardullo}, R. \& {Bond}, H.~E. 1996, \aj, 111, 2332

\bibitem[{{Dreizler} {et~al.}(1995){Dreizler}, {Werner}, \& {Heber}}]{Dre95}
{Dreizler}, S., {Werner}, K., \& {Heber}, U. 1995, Lecture Notes in Physics,
  Berlin Springer Verlag, 443, 160

\bibitem[{{Gautschy}(1997)}]{Gautschy97}
{Gautschy}, A. 1997, \aap, 320, 811

\bibitem[{{Herwig}(2001)}]{Herwig01}
{Herwig}, F. 2001, \apss, 275, 15

\bibitem[{{Miksa} {et~al.}(2002){Miksa}, {Deetjen}, {Dreizler}, {Kruk},
  {Rauch}, \& {Werner}}]{Mik02}
{Miksa}, S., {Deetjen}, J.~L., {Dreizler}, S., {et~al.} 2002, \aap, 389, 953

\bibitem[{{Napiwotzki} \& {Schoenberner}(1991)}]{NapScho91}
{Napiwotzki}, R. \& {Schoenberner}, D. 1991, \aap, 249, L16

\bibitem[{{Quirion} {et~al.}(2004{\natexlab{a}}){Quirion}, {Fontaine}, \&
  {Brassard}}]{QFB04}
{Quirion}, P.-O., {Fontaine}, G., \& {Brassard}, P. 2004{\natexlab{a}}, \apj,
  610, 436

\bibitem[{{Quirion} {et~al.}(2004{\natexlab{b}}){Quirion}, {Fontaine}, \&
  {Brassard}}]{QFByale}
{Quirion}, P.-O., {Fontaine}, G., \& {Brassard}, P. 2004{\natexlab{b}}, in ESA
  SP-559: SOHO 14 Helio- and Asteroseismology: Towards a Golden Future, 598

\bibitem[{{Saio}(1996)}]{Saio96}
{Saio}, H. 1996, in ASP Conf. Ser. 96, Hydrogen-deficient star, ed. C. S.
  Jeffery \& U. Heber (San Francisco: ASP), 361

\bibitem[{{Schuh} {et~al.}(2000){Schuh}, {Dreizler}, {Deetjen}, {Heber}, \&
  {Geckeler}}]{Schu00}
{Schuh}, S., {Dreizler}, S., {Deetjen}, J.~L., {Heber}, U., \& {Geckeler},
  R.~D. 2000, Baltic Astronomy, 9, 395

\bibitem[{{Silvotti}(1996)}]{Silvotti96}
{Silvotti}, R. 1996, \aap, 309, L23

\bibitem[{{Stanghellini} {et~al.}(1991){Stanghellini}, {Cox}, \&
  {Starrfield}}]{Stan91}
{Stanghellini}, L., {Cox}, A.~N., \& {Starrfield}, S. 1991, \apj, 383, 766

\bibitem[{{Starrfield} {et~al.}(1985){Starrfield}, {Cox}, {Kidman}, \&
  {Pensnell}}]{Star85}
{Starrfield}, S., {Cox}, A.~N., {Kidman}, R.~B., \& {Pensnell}, W.~D. 1985,
  \apjl, 293, L23

\bibitem[{{Starrfield} {et~al.}(1984){Starrfield}, {Cox}, {Kidman}, \&
  {Pesnell}}]{Star84}
{Starrfield}, S., {Cox}, A.~N., {Kidman}, R.~B., \& {Pesnell}, W.~D. 1984,
  \apj, 281, 800

\bibitem[{{Starrfield} {et~al.}(1983){Starrfield}, {Cox}, {Hodson}, \&
  {Pesnell}}]{Star83}
{Starrfield}, S.~G., {Cox}, A.~N., {Hodson}, S.~W., \& {Pesnell}, W.~D. 1983,
  \apjl, 268, L27

\bibitem[{{Vauclair} {et~al.}(2005){Vauclair}, {Solheim}, \&
  {{\O}stensen}}]{Vau05}
{Vauclair}, G., {Solheim}, J.-E., \& {{\O}stensen}, R. 2005, \aap, preprint
  doi:101051/0004-6361:20041999

\bibitem[{{Werner}(1992)}]{Werner92}
{Werner}, K. 1992, Lecture Notes in Physics (Berlin: Springer Verlag), 401, 273

\end{thebibliography}

\end{document}